\documentclass[twocolumn,letter]{jpsj2} 

\def\runauthor{H. Fukazawa {\it et al.}}

\title{%
$^{75}$As NMR Study of Hole-Doped Superconductor Ba$_{1-x}$K$_{x}$Fe$_{2}$As$_{2}$ ($T_{\rm c} \simeq 38~$K)
}

\author{%
Hideto~Fukazawa$^{1,2}$\thanks{E-mail address: hideto@nmr.s.chiba-u.ac.jp}, 
Takehiro~Yamazaki$^{1}$,
Kenji~Kondo$^{1}$,
Yoh~Kohori$^{1,2}$, 
Nao~Takeshita$^{2,3}$,
Parasharam~M.~Shirage$^{3}$,
Kunihiro~Kihou$^{3}$,
Kiichi~Miyazawa$^{3,4}$,
Hijiri~Kito$^{2,3}$, 
Hiroshi~Eisaki$^{2,3}$, 
and Akira~Iyo$^{2,3}$
}

\inst{%
$^{1}$Department of Physics, Chiba University, Chiba 263-8522, Japan\\
$^{2}$JST, TRIP, Chiyoda-ku, Tokyo 102-0075, Japan\\
$^{3}$National Institute of Advanced Industrial Science and Technology, Tsukuba 305-8562, Japan\\
$^{4}$Department of Applied Electronics, Tokyo University of Science, Noda 278-8510, Japan
}

\recdate{\today}

\abst{%
We report the $^{75}$As nuclear magnetic resonance (NMR) measurement of the hole-doped superconductor Ba$_{1-x}$K$_{x}$Fe$_{2}$As$_{2}$ 
with different lattice parameters and different superconducting volume fractions ($T_{\rm c} \simeq 38~$K). 
$^{75}$As-NMR spectra revealed that the magnetically ordered and superconducting phases are microscopically separated.  
The spin-lattice relaxation rate $1/T_{1}$ in the normal state reflects the existence of 
a large two-dimensional antiferromagnetic spin fluctuation. 
The $1/T_{1}$ in the superconducting state down to the lowest measurement temperature $T$ varies close to $T^{3}$. 
In addition, it exhibits no coherence peak just below $T_{\rm c}$. 
This shows a $T$ dependence similar to those of other iron pnictides. 
}

\kword{%
Ba$_{1-x}$K$_{x}$Fe$_{2}$As$_{2}$, superconductivity, magnetic order, phase separation, nuclear magnetic resonance
}

\begin{document}
\maketitle


The discovery of superconductivity in F-doped LaFeAsO with a superconducting transition temperature $T_{\rm c}=26$~K~\cite{Kam1} 
has accelerated further investigations of related 
superconductors~\cite{Kit1,Ren2,GCh1,XCh1,Rot2,Hsu1,Wan1,Mat1}. 
The $3d$ electrons originating from an FeAs layer form multiple bands at the Fermi level and 
play an important role in superconductivity
~\cite{Maz1,Kur1}. 
In particular, nondoped materials commonly exhibit an antiferromagnetic (AF) order with an adjacent structural phase transition, 
which resembles the parent materials of high-$T_{\rm c}$ cuprates. 
Hence, the relation between magnetic order and superconductivity is one of the vital issues in the investigations of such compounds. 

K-doped BaFe$_{2}$As$_{2}$ is the firstly reported oxygen-free iron-pnictide superconductor with $T_{\rm c} = 38~$K~\cite{Rot2}. 
The crystal structure of Ba$_{1-x}$K$_{x}$Fe$_{2}$As$_{2}$ is of the ThCr$_{2}$Si$_{2}$-type. 
This structure possesses an FeAs layer similar to that realized in LaFeAsO. 
The parent material BaFe$_{2}$As$_{2}$ exhibits AF anomaly at $T_{\rm N} =$~140~K~\cite{Rot1}. 
Neutron diffraction measurements revealed an ordered moment of 0.87~$\mu_{\rm B}$ (Bohr magneton) at the Fe site 
with a $q$ vector of $(1,0,1)$ for the orthorhombic structure~\cite{Hua1}. 
It is important to note that this compound exhibits structural phase transition 
as well as AF anomaly~\cite{Rot1,Hua1}.  
The zero-field $^{75}$As-NMR spectrum also revealed that the magnetically ordered state of this compound is commensurate~\cite{Fuk2,Fuk3}. 
The evaluated $H_{\rm int}$ at 4.2~K decreases gradually with increasing pressure~\cite{Fuk5}. 
The results are consistent with the other group's NMR results obtained using single crystals~\cite{Ktg1}. 

Several groups reported the phase diagram of Ba$_{1-x}$K$_{x}$Fe$_{2}$As$_{2}$~\cite{Rot3,HCh1}. 
For the compound, the AF anomaly disappears and superconductivity is induced by hole doping, K substitution for Ba. 
The superconducting (SC) state is confirmed to be the bulk from the specific heat jump at $T_{\rm c}$~\cite{Mu1}. 
The most striking feature of the phase diagram is that the SC region widely spreads for $0.2 \leq x \leq 1$ 
and thus the possibility of the coexistence of the AF and SC states exists for $0.2 \leq x \leq 0.4$. 
Bulk measurements cannot determine whether or not this coexistence is macroscopic. 
Hence, measurements with a local probe are strongly required. 
Recent $\mu$SR measurements show the existence of the phase-separated AF and SC states~\cite{Gok1,Par1}. 
Here, we report the $^{75}$As-NMR measurements of Ba$_{1-x}$K$_{x}$Fe$_{2}$As$_{2}$ 
with different lattice parameters and different superconducting volume fractions ($T_{\rm c} \simeq 38~$K) 
in order to investigate the gap symmetry of superconductivity 
and establish the nature of the coexistence of the AF and SC states. 

To date, intensive NMR studies are performed on other iron pnictide superconductors~\cite{Nak1,Mata1,Gra1,Muk1,Kaw1,Nin1,Kot1}. 
Knight shift measurements revealed that the spin-singlet superconductivity is realized.  
The common feature of $1/T_{1}$ in the SC state is the $T^{3}$ behavior at low temperatures 
without $T$ linear behavior, which indicates the absence of the residual density of states. 
This is apparently ascribable to the spin-singlet superconductivity with a nodal structure: $d$-wave superconductivity. 
However, the results of other experimental techniques suggest the fully gapped superconductivity: 
$s$-wave superconductivity~\cite{Mu1,Has1,Naka1}. 
Nagai {\it et al.} calculated $1/T_{1}$ and showed that the anisotropic $s_{\pm}$-wave pairing, 
in which the sign of the SC order parameter changes between Fermi surfaces, explains the apparently controversially experimental results~\cite{Nag1}.  
To the best of our knowledge, this is the first report on the NMR study of hole-doped Fe-based superconductor.


Polycrystalline Ba$_{1-x}$K$_{x}$Fe$_{2}$As$_{2}$ samples were synthesized by a high-pressure method~\cite{Kit1}. 
X-ray diffraction analysis revealed that the obtained samples have different lattice parameters, 
which is associated with the temperature control or the starting ratio of (Ba, K) and FeAs. 
In this study we used two different batches of samples and fixed the ratio of starting Ba to K at 0.6:0.4: 
$a=3.923~{\rm \AA},c=13.220~{\rm \AA}$ sample and $a=3.926~{\rm \AA},c=13.297~{\rm \AA}$ sample. 
According to the relation between the lattice parameter and the nominal composition $x$~\cite{Rot3}, 
the materials correspond to $x=0.3$ and 0.4. 
For convenience, we utilize these $x$'s hereafter in this paper. 
Although the $x=0.3$ sample was confirmed to be of nearly single phase by X-ray diffraction analysis, 
the $x=0.4$ sample contained less than 10\% FeAs impurity. 
We also determined the $T_{\rm c}$ and SC volume fraction of the samples with a commercial SQUID magnetometer. 
The $T_{\rm c}$'s of the samples are about 38~K and  
the volume fractions for $x=0.3$ and 0.4 are 75\% and 50\%, respectively. 
The NMR experiment on the $^{75}$As nucleus ($I=3/2$, $\gamma = 7.292$~MHz/T) was carried out 
using phase-coherent pulsed NMR spectrometers and a superconducting magnet. 
We performed NMR measurement at around 6~T. 
According to the report on the upper critical field of Ba$_{1-x}$K$_{x}$Fe$_{2}$As$_{2}$~\cite{Ni1,ZWan1}, 
the $T_{\rm c}(\sim 6~{\rm T})$ is 35~K for the samples of $T_{\rm c}(0~{\rm T}) = 38$~K. 
The samples were crushed into powder for use in the experiments. 
The powder was oriented at 150~K at a field of 6~T. 
The $c$-axis of each microcrystal in oriented powders is perpendicular to the external field~\cite{Fuk2}. 
The NMR spectra were measured by sweeping the applied fields at a constant resonance frequency. 
The origin of the Knight shift $K=0$ of the $^{75}$As nucleus was determined by the $^{75}$As NMR measurement of GaAs~\cite{Bas1}. 
$T_{1}$ was evaluated by a saturation recovery method.


 \begin{figure}
  \centering
  \includegraphics[width=8cm]{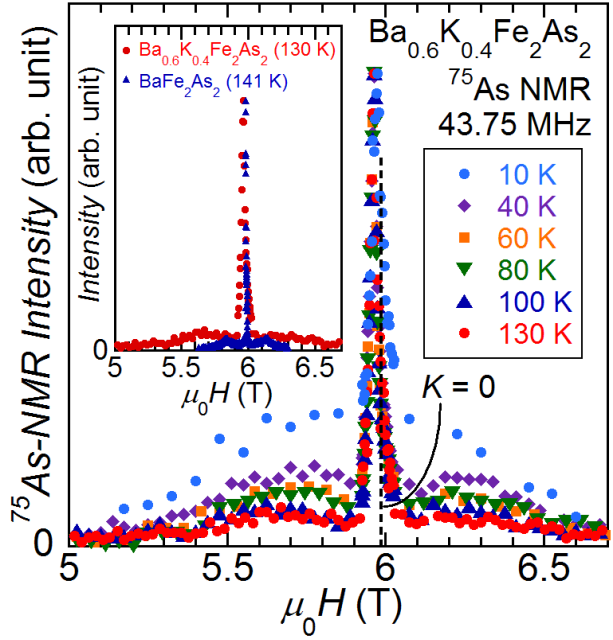}
  \caption{
  (Color online) $^{75}$As NMR spectra of Ba$_{0.6}$K$_{0.4}$Fe$_{2}$As$_{2}$ at various temperatures.
  The inset shows the $^{75}$As NMR spectra of Ba$_{0.6}$K$_{0.4}$Fe$_{2}$As$_{2}$ and BaFe$_{2}$As$_{2}$ in the paramagnetic state. 
  }
 \end{figure}

In Fig.~1, we show the $^{75}$As-NMR spectra of Ba$_{0.6}$K$_{0.4}$Fe$_{2}$As$_{2}$ at various $T$'s. 
Even at higher temperatures, the spectra have broad satellite components in addition to the sharp center line at around 6~T. 
Because K is expected to locate randomly at the (Ba, K) site in Ba$_{0.6}$K$_{0.4}$Fe$_{2}$As$_{2}$, 
these broad satellite components arise from the inhomogeneous distribution of the electric field gradient (EFG) at the As site. 
In the inset of Fig.~1, we show a comparison of the $^{75}$As-NMR spectra of 
Ba$_{0.6}$K$_{0.4}$Fe$_{2}$As$_{2}$ and BaFe$_{2}$As$_{2}$ in the paramagnetic state.
The center line of the spectrum of Ba$_{0.6}$K$_{0.4}$Fe$_{2}$As$_{2}$ is broader than that of BaFe$_{2}$As$_{2}$, 
which is also ascribable to the inhomogeneous distribution of the EFG at the As site and Knight shift. 
Since the EFG at the As site is expected to be parallel to the crystal $c$-axis and thus perpendicular to the external field, 
the $\nu_{Q}$ of Ba$_{0.6}$K$_{0.4}$Fe$_{2}$As$_{2}$ is estimated to distribute at around 5~MHz, 
which is twice larger than the $\nu_{Q}$ of BaFe$_{2}$As$_{2}$. 
The intensity of broad satellite components increases below approximately 80-100~K. 
This is attributable to the AF ordering below $T_{\rm N} = 80$-100~K. 
We will discuss the determination of $T_{\rm N}$ later. 
However, $T_{\rm N}$ is clearly larger than 80~K. 
Hence, this increase in intensity is hardly attributed to the impurity FeAs, which exhibits AF ordering below 77~K~\cite{Sel1}. 
Note that the center line observed in the paramagnetic (PM) state 
coexists even at 10~K with the magnetically ordered component. 
This indicates the occurrence of the phase separation of the AF and PM states at low temperatures, 
which is consistent with the results of $\mu$SR measurements~\cite{Gok1,Par1}. 

 \begin{figure}
  \centering
  \includegraphics[width=8cm]{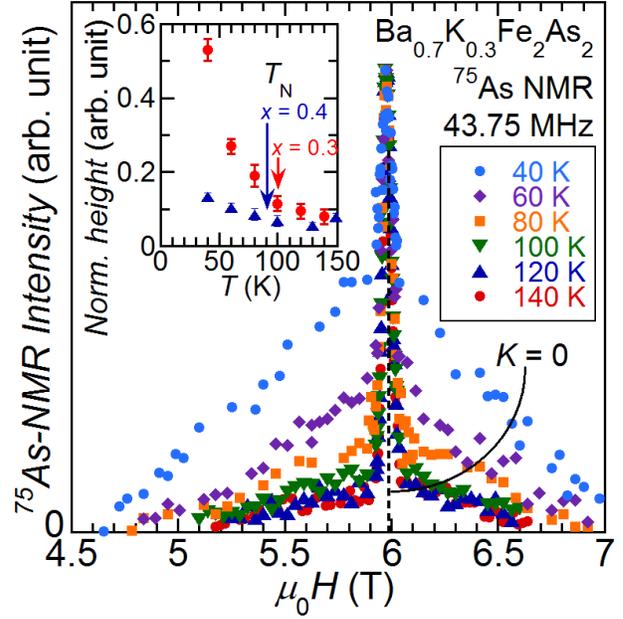}
  \caption{
  (Color online) $^{75}$As NMR spectra of Ba$_{0.7}$K$_{0.3}$Fe$_{2}$As$_{2}$ at various temperatures.
  The inset shows the $T$ dependence of the normalized broad-satellite height 
  against the center-line height in the spectra for $x = 0.3$, 0.4. 
  }
 \end{figure}

In Fig.~2, we show the $^{75}$As-NMR spectra of Ba$_{0.7}$K$_{0.3}$Fe$_{2}$As$_{2}$ at various $T$'s. 
The increase in the intensity of the NMR signals of the satellite components is also observed and 
is much larger than that of Ba$_{0.6}$K$_{0.4}$Fe$_{2}$As$_{2}$. 
In the inset of Fig.~2, we plotted the $T$ dependence of normalized broad-satellite height 
against center-line height for Ba$_{0.7}$K$_{0.3}$Fe$_{2}$As$_{2}$ and Ba$_{0.6}$K$_{0.4}$Fe$_{2}$As$_{2}$ 
in order to estimate AF transition temperature. 
We found that the transition temperature is approximately 100~K for $x=0.3$. 
The temperature is roughly estimated to be between 80 and 100~K for $x=0.4$. 
Hence, AF transition temperature decreases with increasing $x$. 
This is consistent with the macroscopic experimental results of Ba$_{1-x}$K$_{x}$Fe$_{2}$As$_{2}$~\cite{Rot3,HCh1}. 
It is difficult to accurately determine $T_{\rm N}$ for $x=0.4$ 
since the signal intensity of broader satellites for $x=0.4$ is much less than that for $x=0.3$. 
We performed magnetization measurement above $T_{\rm c}$, in order to determine AF transition temperature more accurately. 
However, we could not determine $T_{\rm N}$ because a small amount of ferromagnetic Fe impurity, 
which is not detected by X-ray diffraction analysis, masks the anomaly associated with $T_{\rm N}$. 

We additionally performed zero-external-field NMR measurements at 1.5~K and obtained spectra for $x=0.3$ and 0.4 (not shown data). 
The obtained spectral profiles are basically the same as that of the parent material BaFe$_{2}$As$_{2}$~\cite{Fuk2,Fuk3}. 
By the same analytical procedure described in ref.~\ref{Fuka3}, we evaluated the internal magnetic field $H_{\rm int}$ at the As site 
to be 1.23(5)~T for $x=0.3$ and 1.17(3)~T for $x=0.4$. 
These values are rather smaller than $H_{\rm int}=1.39(2)$~T in BaFe$_{2}$As$_{2}$~\cite{Fuk2} and decrease with $x$. 
Since the magnetic transition temperature is a rough measure of the magnetic order parameter, 
this tendency of $H_{\rm int}$ with $x$ is consistent with 
the smaller $T_{\rm N}$ of these compounds than that of BaFe$_{2}$As$_{2}$. 
Furthermore, we note that there is no consistency between the fraction in the AF region and the SC volume fraction of the samples with $x$. 
At the present stage, we cannot explain this inconsistency. 
The K distribution in the samples might be related to the AF or SC fraction 
since the temperature control condition or the starting ratio of (Ba, K) and FeAs may affect such a local distribution of K. 
However, we cannot answer how K distributes in the AF and SC phases and 
can only state that the samples studied here have identical $T_{\rm c}$'s.

 \begin{figure}
  \centering
  \includegraphics[width=8cm]{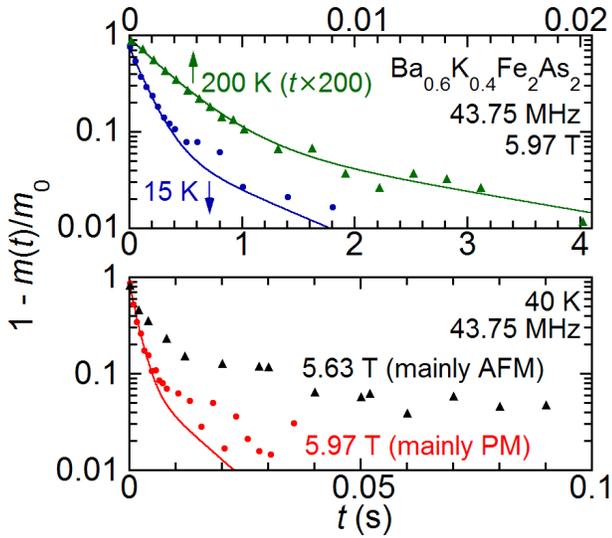}
  \caption{
  (Color online) Nuclear magnetization recovery curves of Ba$_{0.6}$K$_{0.4}$Fe$_{2}$As$_{2}$ at 15, 40, and 200~K. 
  Solid lines denote the fitting curve using the formula written in the text. 
  }
 \end{figure}

In Fig.~3, we show nuclear magnetization recovery curves of Ba$_{0.6}$K$_{0.4}$Fe$_{2}$As$_{2}$ at 15, 40, 200~K. 
We obtained $T_{1}$ at a fixed frequency of 43.75~MHz with an external field of 5.97~T in the $T$ range of 4.2-250~K. 
The nuclear magnetization recovery curve was fitted by the following double-exponential function 
as expected for the center line of the spectrum of the nuclear spin $I=3/2$ of the $^{75}$As nucleus~\cite{Sim1}, 
$$1-\frac{m(t)}{m_{0}} = 0.1\exp\left( -\frac{t}{T_{1}}\right)  + 0.9\exp\left( -\frac{6t}{T_{1}}\right) ,$$
where $m(t)$ and $m_{0}$ are nuclear magnetizations after a time $t$ from the NMR saturation pulse 
and thermal equilibrium magnetization. 
In the paramagnetic state, the data are well fitted to the curve with a single $T_{1}$ component, as clearly seen in the result at 200~K. 
However, the data are less fitted with a single $T_{1}$ component below about 100~K. 
As shown in the data at 15 and 40~K, a deviation of the fitting curve from the experimental data exists and 
a longer $T_{1}$ component appears in the recovery curves. 
This is because the signals at around 5.97~T include signals from the nucleus affected by the AF-ordered state below about 80-100~K, 
which has a longer $T_{1}$ component. 
This is very clear in the data obtained at 5.63~T, at which the signals mainly come from the AF-ordered state. 
Therefore, a shorter $T_{1}$ component is an intrinsic relaxation rate for the center line of the spectrum at around 6~T. 
We tentatively analyzed the recovery curves with double $T_{1}$ components. 
However, we could not obtain a systematic $T$ dependence of $T_{1}$ 
since a longer $T_{1}$ component has worse statistical accuracy than a shorter component. 
Single $T_{1}$ component fitting analysis, which is used to study a shorter $T_{1}$ component, has a systematic $T$ dependence. 
Hence, we used single $T_{1}$ component fitting analysis for all $T$. 

 \begin{figure}
  \centering
  \includegraphics[width=8cm]{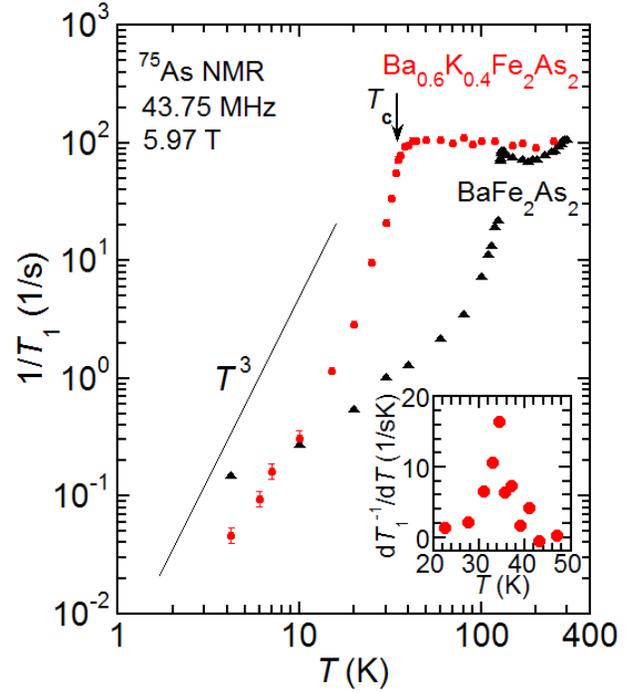}
  \caption{
  (Color online) $1/T_{1}$'s of Ba$_{0.6}$K$_{0.4}$Fe$_{2}$As$_{2}$ and BaFe$_{2}$As$_{2}$. 
  The solid line proportional to $T^{3}$ is a visual guide. 
  The inset shows the derivative of $1/T_{1}$ of Ba$_{0.6}$K$_{0.4}$Fe$_{2}$As$_{2}$.
  }
 \end{figure}

In Fig.~4, we show the $T$ dependence of $1/T_{1}$ of $^{75}$As of Ba$_{0.6}$K$_{0.4}$Fe$_{2}$As$_{2}$ and BaFe$_{2}$As$_{2}$. 
$1/T_{1}$ remains nearly constant above about 40~K in Ba$_{0.6}$K$_{0.4}$Fe$_{2}$As$_{2}$, 
which indicates that the two-dimensional AF spin fluctuation of an itinerant electron system~\cite{Mor1} is predominant in this material. 
Assuming that $1/T_{1}T$ is proportional to a Curie-Weiss-like $T$ dependence $1/(T+\theta_{\rm CW})$ due to AF fluctuation 
in the $T$ range between 40 and 250~K, we obtained the Curie-Weiss temperature $\theta_{\rm CW} =$ -2(2)~K. 
This should be interpreted to indicate that the normal and PM states of Ba$_{0.6}$K$_{0.4}$Fe$_{2}$As$_{2}$ are on the verge of AF instability. 
This $T$ dependence is quite different from that of an electron-doped system~\cite{Sek1} BaFe$_{1.8}$Co$_{0.2}$As$_{2}$~\cite{Nin1}, 
where the Korringa behavior and a pseudogap-like $T$ behavior at higher temperatures are observed, 
and is similar to that of LaFeAsO$_{0.96}$F$_{0.04}$~\cite{Nak1}. 
Since the quantity of the substitution of K for Ba is much larger than that of Co for Fe, 
this result indicates that 3$d$ electronic spin fluctuations are not markedly suppressed by K substitution, that is, by hole doping. 
Moreover, note that the normal state property of Ba$_{0.6}$K$_{0.4}$Fe$_{2}$As$_{2}$ is similar to that of 
underdoped La$_{2-x}$Sr$_{x}$CuO$_{4}$~\cite{Ohs1}. 
Furthermore, the result is consistent with that of the theoretical study of $1/T_{1}$ 
on the basis of fluctuation-exchange approximation, which explains the presence and absence of the pseudogap behavior 
in the $T$ dependence of $1/T_{1}$ of electron-doped and hole-doped iron-pnictide superconductors, respectively~\cite{Ike1}. 

In the inset of Fig.~4, we show the derivative of $1/T_{1}$ of Ba$_{0.6}$K$_{0.4}$Fe$_{2}$As$_{2}$. 
The sharp peak at around 35~K corresponds to the SC transition at a magnetic field of about 6~T. 
Below $T_{\rm c}(5.97~{\rm T})\simeq 35$~K, $1/T_{1}$ rapidly decreases on cooling without a coherence peak. 
At lower temperatures, $1/T_{1}$ varies close to $T^{3}$ and is not exponential; 
the fitting of $1/T_{1}$ below 20~K is proportional to $T^{2.6(3)}$. 
In addition, no $T$ linear behavior was observed down to the lowest measurement temperature of 4.2~K. 
This tendency of $1/T_{1}$ in the SC state is similar to the results commonly reported 
for other iron pnictides~\cite{Nak1,Mata1,Gra1,Muk1,Nin1,Kot1}. 
For Ba$_{1-x}$K$_{x}$Fe$_{2}$As$_{2}$, specific heat~\cite{Mu1} and ARPES~\cite{Naka1} measurements 
suggest a fully gapped superconductivity. 
At the present stage, we cannot exclude the possibility of unconventional superconductivity with a nodal structure. 
However, one of the possible explanations is the anisotropic $s_{\pm}$-wave pairing model~\cite{Nag1}. 
In order to determine the detailed SC gap, direction-sensitive experiments such as specific heat and 
thermal conductivity measurements under accurate magnetic field are strongly required.


In summary, we performed the $^{75}$As nuclear magnetic resonance (NMR) measurement of 
the hole-doped superconductor Ba$_{1-x}$K$_{x}$Fe$_{2}$As$_{2}$ 
with different lattice parameters and different superconducting volume fractions ($T_{\rm c} \simeq 38~$K). 
$^{75}$As-NMR spectra revealed that the magnetically ordered and SC phases are microscopically separated. 
The spin-lattice relaxation rate $1/T_{1}$ in the normal state reflects the existence of 
a large two-dimensional AF spin fluctuation. 
The $1/T_{1}$ in the SC state down to the lowest measurement temperature $T$ varies close to $T^{3}$. 
In addition, it exhibits no coherence peak just below $T_{\rm c}$. 
This shows a $T$ dependence similar to those of other iron pnictides. 
Further studies including those of sample dependence are required to understand the origin of the high $T_{\rm c}$ 
in this category of iron-based superconductors; such studies are in progress.


The authors thank T. Saito and Y. Yamada for their contribution to this work. 
They also thank H. Ikeda for valuable discussion. 
This work is supported by a Grant-in-Aid for Scientific Research from the MEXT, 
Global COE program, AGSST of Chiba University.

\end{document}